\pdfoutput=1
\documentclass[journal]{IEEEtran}

\ifCLASSINFOpdf
  \usepackage[pdftex]{graphicx}
  
\else
  
\fi

\usepackage{microtype}
\usepackage{hyperref}
\usepackage{amsmath}
\usepackage{mathtools}
\usepackage{amssymb}
\usepackage{color,soul}
\usepackage{arydshln}
\usepackage{booktabs} 
\usepackage{algorithmic}
\usepackage[affil-it]{authblk}

\begin{document}
\title{Deep Unfolding with Normalizing Flow Priors for Inverse Problems}
\author[1]{Xinyi Wei}
\author[1]{Hans van Gorp}
\author[1]{Lizeth Gonzalez Carabarin}
\author[2]{Daniel Freedman}
\author[3]{Yonina C. Eldar}
\author[1]{Ruud J.G. van Sloun}

\affil[1]{Department of Electrical Engineering, Eindhoven University of Technology, Eindhoven, The Netherlands}
\affil[2]{Google Research, Tel Aviv, Israel}
\affil[3]{Department of Math and Computer Science, Weizmann Institute of Science, Rehovot, Israel}

\maketitle

\begin{abstract}
Many application domains, spanning from computational photography to medical imaging, require recovery of high-fidelity images from noisy, incomplete or partial/compressed measurements. State-of-the-art methods for solving these inverse problems combine deep learning with iterative model-based solvers, a concept known as deep algorithm unfolding or unrolling. By combining a-priori knowledge of the forward measurement model with learned proximal image-to-image mappings based on deep networks, these methods yield solutions that are both physically feasible (data-consistent) and perceptually plausible (consistent with prior belief). However, current proximal mappings based on (predominantly convolutional) neural networks only implicitly learn such image priors. In this paper, we propose to make these image priors fully explicit by embedding deep generative models in the form of normalizing flows within the unfolded proximal gradient algorithm, and training the entire algorithm end-to-end for a given task. We demonstrate that the proposed method outperforms competitive baselines on various image recovery tasks, spanning from image denoising to inpainting and deblurring, effectively adapting the prior to the restoration task at hand.
\end{abstract}

\begin{IEEEkeywords}
Deep unfolding, normalizing flows, inverse problem, image reconstruction
\end{IEEEkeywords}

\IEEEpeerreviewmaketitle

\section{Introduction}
\label{Introduction}

\begin{figure*}[t!]
    \centering
    \includegraphics[trim=0 440 55 110,clip,width=0.955\textwidth]{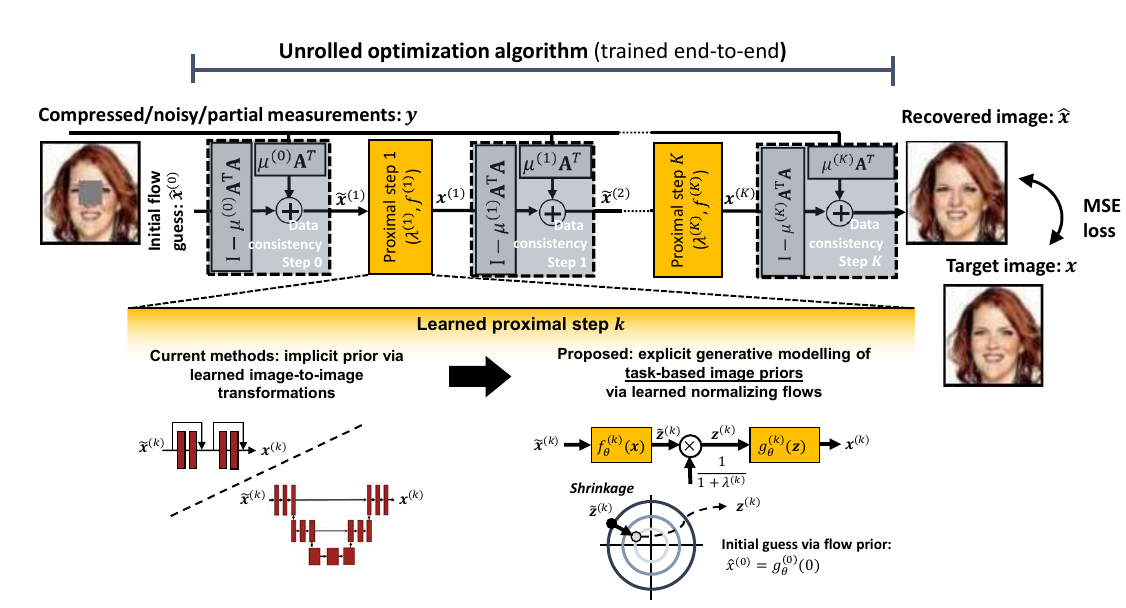}
    \caption{Overview of the proposed algorithm. We unroll the iterations of a proximal gradient algorithm as a deep neural network. The proximal step performs shrinkage in the latent space of a normalizing flow prior, successively pushing the  image likelihood in that distribution across the folds.  \label{fig:overview} }
\end{figure*}

Image reconstruction from noisy, partial or limited-bandwidth measurements is an important problem with applications spanning from fast Magnetic Resonance Imaging (MRI)\cite{pezzotti2020adaptive} to photography \cite{neto2012introduction}. These reconstruction tasks can be posed as linear inverse problems, which are often ill-posed, with many potential solutions satisfying the measurements. Recovery of a meaningful and plausible solution thus requires adequate statistical priors. Formulating such priors for natural or medical image recovery tasks is however not trivial and often dependent on the recovery task itself. 

Traditional convex optimization methods for, e.g., compressed sensing assume sparsity in some transformed domain \cite{donoho2006compressed,eldar2012compressed}. Choosing an appropriate sparse basis is highly dependent on the application and requires careful analysis of, for example, wavelet or total variation-based regularizers that are hard to tune in practice. 

Deep learning \cite{ lecun_deep_2015} is increasingly adopted for image reconstruction, outperforming traditional iterative-based reconstruction methods for tasks such as image denoising \cite{burger12,tian20,vincent08,xie12}, deconvolution \cite{ xu14}, inpainting \cite{ugur18} \cite{ Kamyar19} and end-to-end signal recovery \cite{ kulkarnj16} \cite{ pelt13} \cite{boublil15}\cite{Jin17}.
More specifically, within the framework of compressed sensing, in which signals are to be reconstructed from a set of compressed measurements, deep learning methods have improved both image quality and reconstruction speed \cite{Fang20,bora2017compressed}.

Recent works have shown that, using variable splitting techniques \cite{Boyd11,Deriche03}, any preferred denoiser can be used within classical model-based optimization methods. Typical denoising architectures are based on convolutional autoencoders, U-Nets \cite{Ronneberger15}, or residual networks (ResNets) \cite{He16_resnet}. An interesting special case is DRUNet, having the ability to handle various noise levels via a single model \cite{zhang2020plugandplay}. Related to this, deep generative models (DGMs), such as generative adversarial networks (GANs) \cite{Goodfellow14}, variational autoencoders (VAEs) \cite{kingma2014autoencoding} and normalizing flows \cite{kingma2018glow}, can also serve as meaningful priors for inverse problems in imaging \cite{Hand18,Hand20,bora2017compressed,Asim18,icml2020_2655}. DGMs generate a complex distribution (e.g. that of natural images) from a simple latent base distribution (e.g. independent Gaussians) using a learned deterministic transformation, obtained by pre-training on a large dataset of clean images. This pre-trained model is subsequently used to solve inverse problems by performing gradient-based optimization in their (possibly lower dimensional) latent space. 

While all of these approaches improve upon the hand-crafted sparsity-based priors and exhibit great empirical success, they do not accelerate the optimization process and still rely on time-consuming iterative algorithms. Moreover, their strength, being agnostic of the task and merely concerned with modeling the general image prior, is also a limitation: these approaches do not exploit task-specific statistical properties that can aid the optimization. 

Deep algorithm unfolding aims to address these problems by unrolling the iterative optimization algorithm as a feed forward deep neural network\cite{monga2019algorithm,Gregor10,li2020efficient,Diamond18}. The result is a deep network that takes the structure of the iterations in, for example, proximal-gradient methods, but allows for learning the parameters and/or successive ``neural'' proximal mappings directly from training data \cite{Mardani18}. 
Examples include ADMM-Net, an unfolded version of the iterative ADMM solver \cite{Yang16deep_admm}, ISTA-Net, integrating convolutional networks with sparse-coding-based soft thresholding activations \cite{Zhang18ista}, and CORONA, an unfolded robust PCA algorithm for clutter suppression in Ultrasound \cite{Oren2020}. Other works include D-AMP \cite{Metzler16} \cite{metzler2017learned} inspired by the approximate message passing (AMP) algorithm, neural proximal gradient descent \cite{Mardani18}, and approaches that unfold primal-dual solvers \cite{Wang16PDSM}\cite{Adler18} or half-quadratic splitting methods \cite{Dong18}\cite{Zhang20}.

However, these fast and task-based neural unrolled proximal gradient descent methods no longer explicitly model the underlying statistical priors as a data-generating probability density function. Instead, current methods rely on ``discriminative'' network architectures to model the proximal mapping, such as Resnet- or U-Net-based architectures. Despite their success and vast application space, such implicit priors via neural mappings complicate analysis and steer away from a probabilistic interpretation and modeling of the data distribution.

In this paper, we propose an end-to-end deep algorithm unfolding framework that combines neural proximal gradient descent with generative normalizing flows priors. Our approach first pre-trains a generic flow-based model on natural images by direct likelihood maximization, and subsequently fine-tunes the entire pipeline and priors to adapt to specific image reconstruction tasks. 

Our main contributions are: 
\begin{itemize}
\item We propose a new framework for solving linear inverse problems based on deep algorithm unfolding and normalizing flows priors that adapt to the data and task.
\item We leverage the generative probabilistic nature of our model to yield a strong initial guess: the maximum likelihood solution of the learned flow prior. 
\item We demonstrate strong performance gains over state-of-the-art neural proximal gradient descent baselines on a wide range of image restoration problems. 
\end{itemize}

The remainder of this paper is organized as follows. In Section~\ref{sec:problem}, we introduce the problem and objective. Then, in Section~\ref{sec:flow}, we present our method by first describing generative flow priors in a generic optimization problem, and then proceeding by unrolling the iterations of a proximal gradient algorithm for that optimization problem. In Section~\ref{sec:flowarchitecture} we turn to describing the specific flow architecture we adopt, and continue by detailing the experimental setup in Section~\ref{sec:setup}. The results and specifics for each of the tasks are given in Section~\ref{sec:results}. Finally, in Section~\ref{sec:discussion} we discuss our results and reflect on our approach, and conclude in Section~\ref{sec:conclusion}. 

\begin{figure*}[t!]
    \centering
    \includegraphics[trim=70 310 95 90,clip,width=560px]{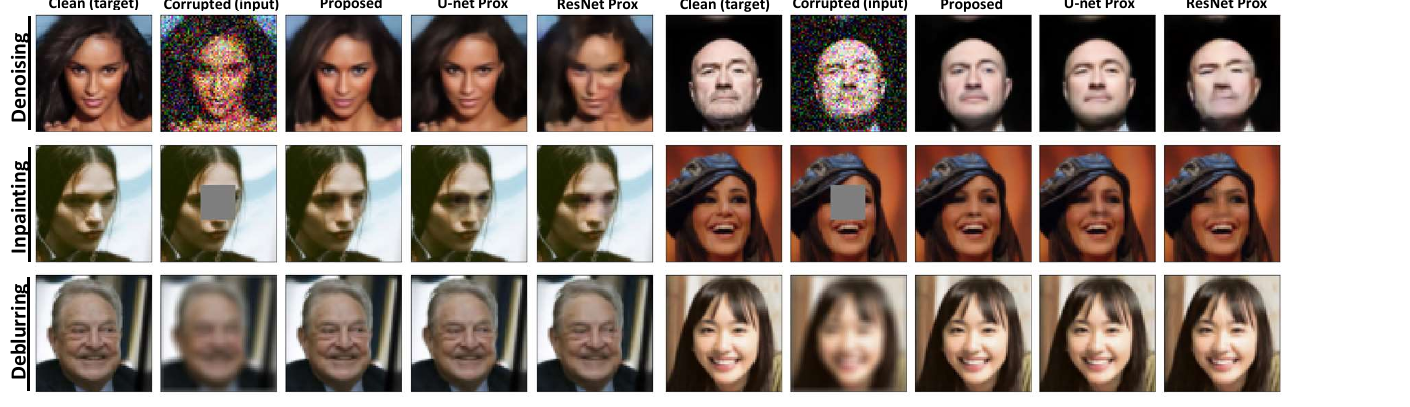}
    \caption{Comparison of the proposed normalizing flows proximal mapping with baselines based on standard U-Net or ResNet proximal mappings across the in-distribution dataset. Results are shown for 3 image restoration tasks applied to 6 typical in-distribution example images.}
    \label{fig:main_results}
\end{figure*}


\section{Problem formulation}
\label{sec:problem}
We consider problems of the following form:
\begin{equation}
\mathrm{y}=\textbf{A}\mathrm{x}+\mathrm{\eta},
\end{equation}
where $\mathrm{y} \in \mathbb{R}^m$ is a noisy measurement vector, $\mathrm{x} \in \mathbb{R}^n$ is the desired signal/image expressed in vector form, $\eta\in \mathbb{R}^m$ is a noise vector, and $\textbf{A} \in \mathbb{R}^{m\times n} $ is a measurement matrix, which we here assume to be known. Note that this can easily be generalized to scenarios where $\textbf{A}$ becomes learnable in the unfolded structure. For the ill-posed inverse problems that we are interested in, maximum-likelihood estimation, i.e., $\mathrm{argmax}_\mathrm{x} p(\mathrm{y}|\mathrm{x})$, is insufficient, yielding solutions that adhere to the measurement model but are visually implausible. By imposing statistical priors,  meaningful solutions (i.e., those that fit expected behavior and prior knowledge) can be obtained through maximum a posteriori (MAP) inference:  
\begin{equation}\label{eq:2}
\hat{\mathrm{x}}_{MAP} \coloneqq \underset{\mathrm{x}}{\arg\max}  p(\mathrm{x}|\mathrm{y})\propto \underset{\mathrm{x}}{\arg\max}  p(\mathrm{y}|\mathrm{x})\,p_{\theta}(\mathrm{x}),
\end{equation}
where $p(\mathrm{y}|\mathrm{x})$ is the likelihood according to the linear noisy measurement model, and $p_{\theta}(\mathrm{x})$ is the signal prior. 

The effectiveness of MAP inference strongly depends on the adequacy of the chosen prior. Formalizing such knowledge is challenging and requires careful analysis for each domain, for example natural images or medical images, and recovery task.
Assuming a Gaussian distribution of measurement model errors, i.e. $p(\mathrm{y}|\mathrm{x}) \sim \mathcal{N}(\mu=\textbf{A}\mathrm{x},\sigma_n^2)$, MAP optimization leads to the following (negative log posterior) minimization problem:
\begin{equation}
\hat{\mathrm{x}} = \underset{\mathrm{x}}{\arg\min}\frac{1}{2\sigma_n^2}\| \mathrm{y}-\textbf{A}\mathrm{x}\|_2^2 - \log p_\theta(\mathrm{x}).
\label{eqn:map}
\end{equation}
Given \eqref{eqn:map}, our goal is now twofold: 1) to learn a useful prior $p_\theta(\mathrm{x})$, and 2) to accelerate and improve performance of standard gradient-based optimization of \eqref{eqn:map} using deep algorithm unfolding. We will address the former in the next section, and then proceed with the latter in Section~\ref{sec:prox_grad}.

\section{Deep algorithm unfolding with flow priors}
\label{sec:flow}
\subsection{Flow priors}
Normalizing flows \cite{kingma2018glow} are a class of generative models, which are capable of modeling powerful and expressive priors. Normalizing flows transform a base probability distribution \(p({\mathrm{z}})\sim \mathcal{N}(0,I)\) into a more complex, possibly multi-modal distribution by a series of composable, bijective, and differentiable mappings. 

Due to the invertible nature of normalizing flows models, they can operate in both directions. These are the generative direction, which generates an image from a point in latent space (\(\mathrm{x} = g_\theta(\mathrm{z})\)), and the flow direction, which maps images into its latent representation (\(\mathrm{z} = f_\theta(\mathrm{x})\)). To create a normalizing flow model of sufficient capacity, many layers of bijective functions can be composed together:
\begin{equation}
    \begin{aligned}
    \mathrm{z} &= f_\theta(\mathrm{x}) =  (\mathrm{f_{1}}\circ\mathrm{f_{2}}\circ...\circ\mathrm{f_{i}})(\mathrm{x}),
    \end{aligned}
\end{equation}
and, 
\begin{equation}
    \begin{aligned}
    \mathrm{x} &= g_\theta(\mathrm{z}) =  (\mathrm{f_{i}^{-1}}\circ\mathrm{f_{i-1}^{-1}}\circ...\circ\mathrm{f_{1}^{-1}})(\mathrm{z}),
    \end{aligned}
\end{equation}
where `\(\circ\)' denotes the composition of two functions, and \(\theta\) are the parameters of the model. 

The functions $f_{1}, f_{2},...f_{i}$ are all layers of a neural network. Each of these layers is invertible and may have trainable parameters. Specifically, each layer is either an actnorm, squeeze, invertible convolution, or affine transformation. Through this lens, $f_{\theta}$ is in fact a neural network of which the inverse also exists. Details of the specific normalizing flow model we adopt, and its parameterization, are given in Section~\ref{sec:flowarchitecture}. 

Exact density evaluation of \(p_{\theta}(\mathrm{x})\) is possible through the use of the change of variables formula, leading to:
\begin{equation}
    \label{eqn:LL}    
    \begin{aligned}
    \log p_{\theta}(\mathrm{x})&=\log p(\mathrm{z})+\log \mid \texttt{det}Df_\theta(\mathrm{x}) \mid,
    \end{aligned}
\end{equation}
where \(D\) is the Jacobian, accounting for the change in density caused by the transformation \(f_\theta\). 

With latent $\mathrm{z}$ following a normal distribution with zero mean and standard deviation, and $\mathrm{x}=g_\theta(\mathrm{z})$, we can then perform proximal updates in $\mathrm{z}$ space:  
\begin{equation}
\begin{split}
\hat{\mathrm{z}} & = \underset{\mathrm{z}}{\arg\min} \frac{1}{2\sigma_n^2} \| \mathrm{y}-\textbf{A}g_\theta(\mathrm{z})\|_2^2 - \log p(\mathrm{z}),
\\
 & = \underset{\mathrm{z}}{\arg\min} \| \mathrm{y}-\textbf{A}g_\theta(\mathrm{z})\|_2^2 +\lambda \|\mathrm{z}\|_2^2,
\end{split}
\label{eqn:glowprior}
\end{equation}
where $\lambda$ is a parameter that balances the importance of adhering to the measurements (data consistency) and the prior. Note that other simple base priors for $\mathrm{z}$ may be adopted as well. For instance, one may assume a Laplace distribution, recasting the $\ell_2$ norm in \eqref{eqn:glowprior} into an $\ell_1$ norm. 

In our case, we have formulated the problem as  \eqref{eqn:glowprior}, finding the optimal latent code $\mathrm{z}$, instead of finding the optimal reconstructions $\mathrm{x}$, as \eqref{eqn:map}. This is useful as both previous works \cite{asim2020invertible} \cite{pmlr-v139-whang21a}, and our own experiments have shown that optimizing in pixel domain is practically much more challenging than in latent domain. Because the statistics in $\mathrm{z}$ space are the same in all dimensions (standard Normal distribution) and semantically useful relationships are created by the normalizing flow network when going to and from the latent space.

\begin{figure*}[t!]
    \centering
    \includegraphics[trim=0 280 95 0,clip,width=560px]{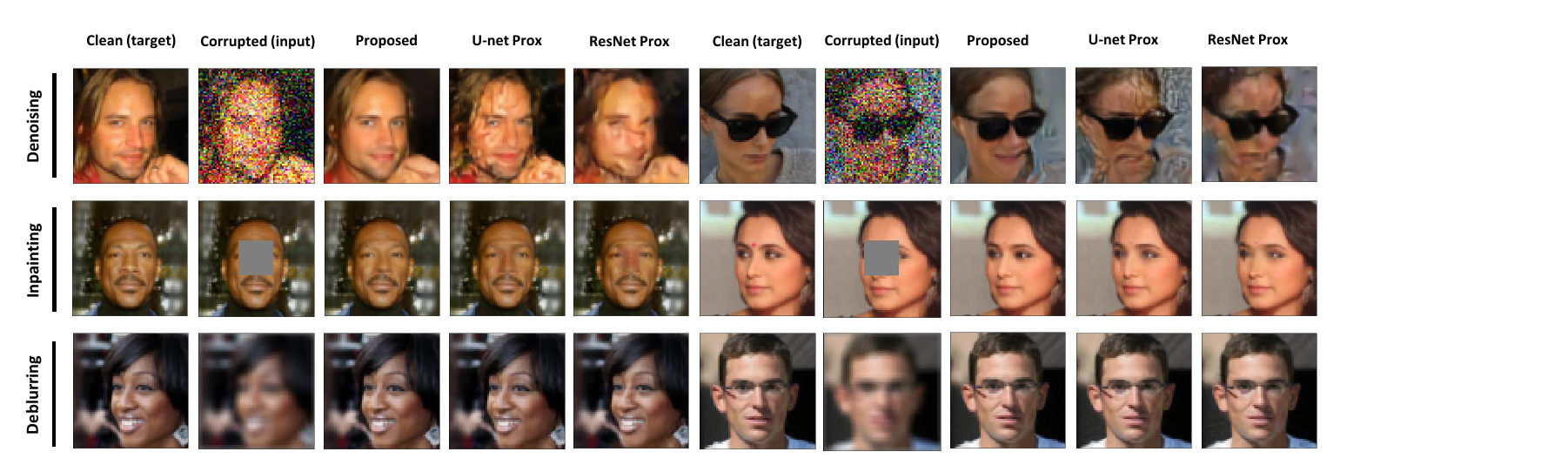}
    \caption{Comparison of the proposed normalizing flows proximal mapping with baselines based on standard U-Net or ResNet proximal mappings on the CelebA-HQ images with out-of-distribution measurements. Results are shown for the denoising, inpainting and deblurring tasks applied to 6 typical CelebA-HQ example images.}
    \label{fig:out_measurement}
\end{figure*}

\subsection{Unrolled proximal gradient iterations}
\label{sec:prox_grad}
The optimization problem in \eqref{eqn:glowprior} can be solved using an iterative proximal-style algorithm that alternates between gradient updates in the direction of the data consistency term and pushing the solution in the proximity of the prior. 

To derive our iterative scheme, we will alternate using the problem formulations in \eqref{eqn:map} and \eqref{eqn:glowprior}.  More specifically, at every fold $k$ the network performs data consistency updates using the $x$-space formulation in \eqref{eqn:map}:
\begin{equation}
    \tilde{\mathrm{x}}^{(k+1)} = \mathrm{x}^{(k)} - \mu^{(k)} \mathbf{A}^T(\mathrm{y} - \mathbf{A}\mathrm{x}^{(k)})
    \label{eqn:dataconsistency}
\end{equation}
where superscript $(k)$ denotes the current fold and $\mu^{(k)}$ is the trainable step size.  The signal representation is then converted to the latent space:
\begin{equation}
    \tilde{z}^{(k+1)} = f_\theta^{(k+1)}(\tilde{\mathrm{x}}^{(k+1)}).
\end{equation}
The purpose of this conversion is so that we may perform the proximal update $\mathcal{P}(\cdot)$ using the $z$-space formulation in \eqref{eqn:glowprior}:
\begin{equation}
    z^{(k+1)} = \mathcal{P}^{(k+1)}(\tilde{\mathrm{z}}^{(k+1)}) = \frac{\tilde{\mathrm{z}}^{(k+1)}}{1 + \lambda^{(k+1)}}
\end{equation}
where $\lambda^{(k+1)}$ is a trainable shrinkage parameter. Intuitively, this can be understood as pushing solutions into a high likelihood regime (i.e. closer to the origin in $\mathrm{z}$). Finally, we convert from latent space back to signal space
\begin{equation}
    \mathrm{x}^{(k+1)} = g_\theta^{(k+1)}(\mathrm{z}^{(k+1)})
\end{equation}
and continue on to the next iteration. 

We unfold the above iterative algorithm as a $K$-fold feedforward neural network that we train end-to-end. After $K$ folds the final estimate \(\hat{\mathrm{x}}\) is produced from the latent space after data consistency:
\begin{equation}
    \begin{aligned}
    \hat{\mathrm{x}} = g_\theta^{(K)}(\Tilde{\mathrm{z}}^{(K)}).
    \end{aligned}
    \label{eqn:finalstep}
\end{equation}

\subsection{End-to-end task-adaptive training and initial guess}
We make use of a pre-trained single generative normalizing flow model from a set of clean images to learn a generic density function. After pre-training, we embed these generic priors into the unrolled architecture and tailor it to the specific recovery task at hand using end-to-end supervised learning. This yields an architecture in which each fold has a distinct and task-based flow model. By unrolling and training end-to-end we no longer guarantee nor explicitly promote normality of the latent space of the flow prox at each fold. We will further discuss this and its implications in Section~\ref{sec:discussion}.

We also make use of the explicit likelihood modeling of the normalizing flow prior to yield a useful initial guess for $x$: the maximum likelihood solution according to the clean images, by exploiting the fact that the most likely image is at $z=0$. This serves as an input to our network and is denoted by $\hat{x}^{(0)}=g_\theta^{(0)}(z^{(0)}=0)$; see also Fig.~\ref{fig:overview}. 

\begin{table*}[t!]
\caption{Experimental results of deep unfolding with normalizing flows priors compared to two strong baselines. Values reported are mean Peak Signal to Noise Ratio (PSNR) of the reconstructed images across the test set.
\label{table:1}}
\centering
\begin{tabular}{lllll}

\toprule

\textbf{Experiment (settings)} & \textbf{Denoising} 
& \textbf{Inpainting} & \textbf{Deblurring}\\

& $\eta \sim \mathcal{N}(\mu_n=0,\,\sigma_n=0.2)$
& $W=19 \times 19$  & $\sigma_b = 5$\\
\toprule

\textbf{ResNet Prox} \cite{Mardani18}& 
27.054 dB & 34.010 dB &
35.554 dB \\
\hdashline
\textbf{Pre-trained ResNet prox} & & 
33.216 dB & 35.703 dB \\
*Pre-trained as a denoiser &&& \\

\toprule

\textbf{U-Net Prox } \cite{ronneberger2015u}& 
28.180 dB& 35.157 dB&
38.662 dB\\
\hdashline
\textbf{Pre-trained U-Net prox} & & 
35.337 dB & 38.046 dB\\
*Pre-trained as a denoiser &&&\\

\toprule

\textbf{Pre-trained GLOW prox} (\textbf{Proposed})&\textbf{28.236} dB & \textbf{35.927} dB &\textbf{40.549} dB\\
*Pre-trained by data-likelihood maximization & & & \\

\toprule
\end{tabular}
\end{table*}

\section{Normalizing Flows architecture}
\label{sec:flowarchitecture}
For the normalizing flow model we make use of GLOW \cite{kingma2018glow}, a normalizing flow architecture that uses \(1\times1\) convolutions to permute the dimensions on a multi-scale architecture \cite{Dinh16}. Central to GLOW is the affine transformation, which is defined as:


\begin{equation}
    \begin{aligned}
    \mathrm{y} = s\cdot\mathrm{x} + t,
    \end{aligned}
    \label{eqn:affine}
\end{equation}
where \(s\) is the scale, and \(t\) is the translation that is applied to \(\mathrm{x}\).
In general, each step of GLOW consists of three stages: actnorm,  an invertible \(1\times1\) convolution, and an affine coupling layer. The actnorm stage (short for activation normalization) is a normalization scheme that is better adapted to low batch sizes than conventional batch normalization. The actnorm stage is implemented as an affine transformation that acts upon the incoming data, with learnable scale and translation parameters.

After normalization an invertible \(1\times1\) convolution is performed. This convolution can be viewed as a generalization of a permutation operation. Permutation of the dimensions is a vital step in normalizing flows to ensure that each dimension can affect every other dimension, enabling the model to transform complex (data) distributions into normal distributions. By learning this permutation as a convolution, rather than choosing it \textit{a-priori}, the permutation can more powerfully adapt to the data and/or task.

Each step of GLOW is concluded with an affine coupling layer. The scale and translation parameters are created from part of the incoming tensor, using:
\begin{gather*}
    \mathrm{x}_a,\mathrm{x}_b = \text{split}(\mathrm{x})\\
    (\log s, t) = \text{NN}(\mathrm{x}_b) \\
    s = \exp(\log s) \\
    \mathrm{y}_a = s\cdot\mathrm{x}_a + t \\
    \mathrm{y}_b = \mathrm{x}_b \\
    \mathrm{y} = \text{concat}(\mathrm{y}_a,\mathrm{y}_b),
\end{gather*}
where NN is a neural network that is not required to be invertible. This is because its input is left unchanged by the affine coupling layer, it changes $x_a$ into $y_a$ via a scaling and transformation, but leaves $x_b = y_b$. Thus when going backwards through this layer, we will always know what input was used for this neural network to generate $s$ and $t$. In order for the network to also affect the '$b$' part of the input, the aforementioned \(1\times1\) convolutions are used, which change which parts of the signal are $a$ and which parts are $b$, thereby slowly allowing GLOW to whiten the entire signal.

All of these steps are combined within a multi-scale architecture \cite{Dinh16} having 6 levels and a depth of 32. For details regarding its architecture and implementations (code) we refer the reader to the original paper by Kingma and Dhariwal \cite{kingma2018glow} as well as the paper by Asim \textit{et al}. \cite{icml2020_2655}.

\section{Experimental setup}
\label{sec:setup}
We assess our method's performance for various image recovery tasks using both the in-distribution images as CelebA-HQ\cite{Karras18}, and the out-of-distribution images as the Anime Faces set \cite{rakshit2020animefaces}. Across all experiments we used \(K=4\) folds of the unfolded proximal gradient algorithm.

\subsection{Dataset}
As in-distribution set we make use of the CelebA-HQ training set. CelebA-HQ consists of 23,000 training, 1,500 validation and 1,500 test images, which are cropped to the faces, and resized to a size of $64 \times 64$ pixels with 3 color channels.

As out-distribution set we make use of the Anime Faces set \cite{rakshit2020animefaces}. As this set was only used for out-of-distribution evaluation performance, no train-test split was made. Similar to the CelebA-HQ dataset the faces were detected  and centered and the images were then cropped to a size of $64 \times 64$ pixels with 3 color channels.

\subsection{Training strategy}
\label{sec:training}
We make full use of the generative capabilities of our GLOW prox, by first pre-training it as a generative model, and consequently re-training it as the proximal operator in an unrolled proximal gradient scheme. 

The pre-trained GLOW architecture consists of a sequence of affine transformations with a depth of flow 18. The number of multi-scale levels is 4. The model is trained on maximizing the data-likelihood of clean images given by:
\begin{equation}
    \begin{aligned}
   \log p_{\theta} (\mathcal{D}) =  \sum_{i=1}^N \left[ \log p_z(f_\theta(\mathrm{x}_i)) + \log\mid \texttt{det} Df_\theta(\mathrm{x}_i) \mid \right].
    \end{aligned}
    \label{eqn:pretrainloss}
\end{equation}

After pre-training, we untie the weights of the GLOW model at every fold and use it as the proximal operator. We then train the proximal gradient network again using the CelebA-HQ dataset, but this using corrupted-clean image pairs. We train the proximal gradient network for the specific image recovery tasks using a Mean Square Error (MSE) loss. We employ the Adam optimizer with (lr = \(1e^{-5}\), \(\beta_1 = 0.9\), \(\beta_2 = 0.999\), and \(\epsilon = 1e-8 \)). Moreover, we train the learnable step size $\mu^{(k)}$, and shrinkage factor $\lambda^{(k)}$ with a higher learning rate, namely \(1e^{-2}\). As before, we perform early stopping based on the validation loss. 

Leveraging the invertible nature of the GLOW model, we strongly reduce train-time memory of the full unfolded architecture using the approach by Putzky and Welling \cite{pputzky2019}; =instead of storing all intermediate activations for back-propagation, we recalculate the gradients given the post-activations during backpropagation.

\subsection{Baselines}
\label{sec:res_denoising}


We compare our generative flow-based priors to two alternative neural proximal mappings; one based on ResNets \cite{Mardani18} and one based on a U-Net. 
Note that for fair and direct comparison, we focus on typical alternatives \textit{within} the unfolded proximal gradient framework. This allows for straightforward assessment of the merit of the proposed (task-adapted) normalizing flow prior, beyond the architectural advantages of unfolding the proximal gradient algorithm itself. Training settings and strategy are identical to those described in Section~\ref{sec:training}. 

Our ResNet proximal baseline follows the structure proposed by Mardani \textit{et al.}, \cite{Mardani18}. Each residual block consists of two convolutional layers ($3\times3$ kernel and 128 feature maps), followed by batch normalization and ReLU activation. This is then followed by another 3 convolutional layers with $1\times1$ kernels. Mardani \textit{et al.} found that using 2 such residual blocks per proximal fold works best, so we follow that here as well. 
Our second baseline proximal mapping is a standard U-Net \cite{ronneberger2015u} implementation. The U-Net is a fully convolutional neural network that consists of a contracting path and an expansive path with skip connections between the two. This allows for learning both low-level and high-level features. We here make use of the \href{https://github.com/usuyama/pytorch-unet.git}{Pytorch U-Net} implementation.

\section{Results}
\label{sec:results}

\subsection{Experiment on in-distribution test images}
\label{sec:in-distribution}

\subsubsection{Denoising}
We consider noisy measurements $\mathrm{y}=\mathrm{x}+\mathrm{\eta}$, where $\mathrm{\eta}$ is an i.i.d. Gaussian noise vector with standard deviation $\sigma_n$ and mean $\mu_n=0$. Note that $\textbf{A}$ in \eqref{eqn:dataconsistency} is thus an identity matrix here. We train the networks on denoising level of $\sigma_n=0.2$ and analyze performance accordingly. The goal of the recovery algorithm is to denoise the image, recovering $\mathrm{x}$ from $\mathrm{y}$.

Table~\ref{table:1} shows that the proposed GLOW prox outperforms the baselines. Qualitatively, the examples displayed in Fig.~\ref{fig:main_results} (top row) show that both the GLOW prox and U-Net prox well preserve the desired features (e.g. the wrinkles around the cheeks).

\subsubsection{Inpainting}
We perform measurements $\mathrm{y}=\textbf{A}\mathrm{x}$, where $\textbf{A}$ is a matrix masking the center $W=19\times19$ elements of an image by operating on its vectorized form $\mathrm{x}$. The goal of the recovery algorithm is to ``inpaint'' the masked pixels as accurately as possible. 

For inpainting, the performance gain using GLOW prox is about 0.6 dB (see table~\ref{table:1}). Visual inspection of the second inpainting example given in the second row of Fig.~\ref{fig:main_results} shows that the proposed method is better capable of producing high-resolution reconstructions. Note that this apparent higher fidelity is paired with a pixel-wise PSNR improvement: reconstructions are not merely visually pleasing, but also more accurate. We again refer to the challenging examples (row 2): while U-Net prox also can reconstruct the actual shape of the nose, the proposed method does produce improved skin tone compared to the surroundings of the "inpainted" area.   

\subsubsection{Deblurring}
We take measurements $\mathrm{y}=\textbf{A}\mathrm{x}$, where $\textbf{A}$ is a 2D convolution matrix blurring the image with a Gaussian kernel having standard deviation $\sigma_b=5$~pixels. We analyze the performance of our method in recovering the original image from the blurred measurements, i.e. deblurring.  

As for the other tasks, the proposed GLOW prox outperforms the other baselines (see table~\ref{table:1}). While the performance gain is about 1.8 dB PSNR with respect to the ResNet prox and the U-Net prox, minor differences could be visually inspected from Fig.~\ref{fig:main_results} (bottom row).

\subsubsection{Impact of pre-training}
\label{sec:impactofpretraining}

We also pre-train the baselines as the denoisers for a fair comparison and evaluate their performances accordingly. Specifically, we first pre-train the baselines in a denoising way using the CelebA-HQ training set. We then embed the denoisers in the unfolded network to train on the image inpainting and deblurring tasks, respectively. Table~\ref{table:1} indicates that U-Net prox does slightly benefit from such pre-training on the inpainting task, and the pre-trained ResNet prox also has a performance gain on the deblurring task. However, they still do not outperform our proposed method.

\subsubsection{Comparison to the plug-and-play approach}
The plug-and-play approach is also a valuable baseline for evaluation. Specifically, we plug the pre-trained GLOW into the unfolded framework and only adapt the step sizes to the specific task in the train time. We observe a significant performance gain (see table~\ref{table:2}) of our proposed method to the plug-and-play approach in all the tasks, further confirming the importance of adapting GLOW at each fold to specific tasks.

\subsection{Experiment on out-of-distribution measurement}
We again test our proposed network and the baselines on the denoising, inpainting, and deblurring tasks using the CelebA-HQ test set, while with out-of-distribution measurement matrix $\textbf{A}$ or noise level $\mathrm{\eta}$. Consequently, for denoising, we test on the noise level $\mathrm{\eta}$ following a standard deviation $\sigma_n=0.25$ and mean $\mu_n=0$. Next, we test the images with the data consistency matrix $\textbf{A}$ at the last fold blocking the center $W=16\times16$ elements for inpainting. Finally, we test the images blurred with a Gaussian kernel $\textbf{A}$ with a standard deviation $\sigma_b=8$~pixels for deblurring. 

Table~\ref{table:3} indicates that our GLOW prox achieves performance gains in all the cases. The visual examples in the first row of Fig.~\ref{fig:out_measurement} strongly confirm that the reconstructions of our method are much sharper and more capable of preserving details (such as hairs). The second row of Fig.~\ref{fig:out_measurement} again proves that the reconstructions of our method are more accurate, given the well-matched skin tone and shape of the front/side noses of the masked area.

\subsection{Experiment on out-of distribution test images}
Finally, we evaluate the behaviour of the proposed method on out-of-distribution data $\mathrm{x}$. To that end, we test our models trained on the CelebA-HQ dataset on images from the Anime Faces set \cite{rakshit2020animefaces}. The images in this out-of-distribution set show clear differences with respect to the train set, such as much bigger eyes and much smaller noses. We evaluate performance on 100 images that were randomly selected from the full dataset. 

\begin{figure*}[t!]
    \centering
    \includegraphics[trim=70 310 95 90,clip,width=560px]{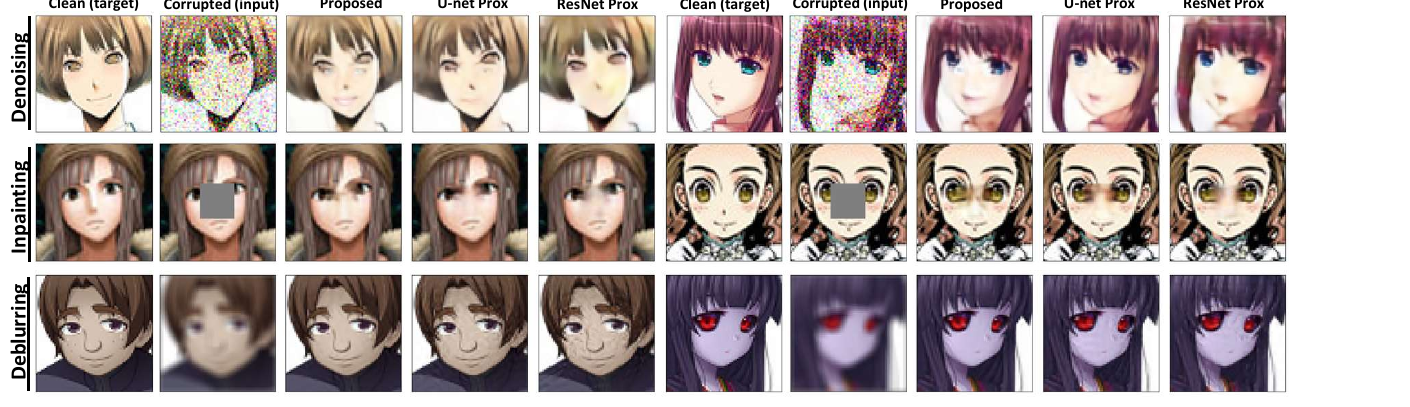}
    \caption{Comparison of the proposed normalizing flows proximal mapping with baselines based on standard U-Net or ResNet proximal mappings. Results are shown for the denoising, inpainting and deblurring tasks applied to 6 typical out-of-distribution example images.}
    \label{fig:out}
\end{figure*}

Table~\ref{Table:4} displays that performance deteriorates for out-of-distribution predictions. U-Net prox outperforms our model by around 0.5 dB in both the image denoising and inpainting tasks, and our model achieves about 1dB gain to the baselines in the deblurring task. 
Example images for visual assessment are given in Fig.~\ref{fig:out}. Interestingly, visual comparison qualitatively shows a stronger ``CelebA'' image prior in the reconstructions by the proposed model compared to the reconstructions of the baselines. This is particularly evident form the inpainting task, where natural noses and eyes are painted into the unnatural Anime faces. In this case the U-Net and the ResNet-based reconstructions are more smooth and blurry, with a less clear ``CelebA fingerprint''.

\begin{table*}[t!]
\caption{Results of deep unfolding with normalizing flows priors compared between with proposed framework and the plug-and-play. Values reported are mean Peak Signal to Noise Ratio (PSNR) of the reconstructed images across the in-distribution test set.  
\label{table:2}}
\centering
\begin{tabular}{lllll}
\toprule
\textbf{Experiment (settings)} & \textbf{GLOW Prox}  & \textbf{GLOW Prox}  &\\
& (Plug-and-play)  & (Proposed)  &\\
\toprule

Denoising ($\eta \sim \mathcal{N}(\mu_n=0,\,\sigma_n=0.2)$)          
&  8.112 dB             & \textbf{28.236} dB                   &\\

Inpainting ($W=19 \times 19$) 
& 18.960 dB             & \textbf{35.927} dB                   &\\

Deblurring ($\sigma_b = 5$)
& 23.141 dB             & \textbf{40.549} dB                   &\\
\toprule
\end{tabular}
\end{table*}


\begin{table*}[t!]
\caption{Experimental results of deep unfolding with normalizing flows priors compared to two strong baselines on the CelebA-HQ images with out-of-distribution measurements. Values reported are mean Peak Signal to Noise Ratio (PSNR) of the reconstructed images across the test set.  
\label{table:3}}
\centering
\begin{tabular}{lllll}
\toprule
\textbf{Experiment (settings)} & \textbf{ResNet Prox } & \textbf{U-Net Prox }  & \textbf{GLOW Prox} &  \\
& \cite{Mardani18} & \cite{ronneberger2015u} & (Proposed) \\
\toprule
Denoising ($\eta \sim \mathcal{N}(\mu_n=0,\,\sigma_n=0.25)$)          
& 25.489  dB            & 25.770 dB             & \textbf{26.633} dB \\
Inpainting ($W=16 \times 16$)          
& 34.644 dB             & 35.917 dB             & \textbf{36.634} dB   \\
Deblurring ($\sigma_b = 8$)         
& 35.358 dB             & 38.034dB             & \textbf{39.461} dB   \\

\toprule
\end{tabular}
\end{table*}

\begin{table*}[t!]
\caption{Experimental results of deep unfolding with normalizing flows priors compared to two baselines. Values reported are mean Peak Signal to Noise Ratio (PSNR) of the reconstructed images across an out-of-distribution test set.  
\label{Table:4}}
\centering
\begin{tabular}{lllll}
\toprule
\textbf{Experiment (settings)} & \textbf{ResNet Prox } & \textbf{U-Net Prox }  & \textbf{GLOW Prox} &  \\
& \cite{Mardani18} & \cite{ronneberger2015u} & (Proposed) & \\
\toprule

Denoising ($\eta \sim \mathcal{N}(\mu_n=0,\,\sigma_n=0.2)$ )
& 21.746 dB           & \textbf{22.587} dB      
& 22.102 dB                      &  \\

Inpainting ($W=19 \times 19$ )
& 23.705 dB            & \textbf{24.015} dB     
& 23.472 dB                    &  \\
                                     
Deblurring ($\sigma_b = 5$)
& 25.007 dB            & 26.224 dB             & \textbf{27.419} dB                      &  \\

\toprule
\end{tabular}
\end{table*}

\section{Discussion}
\label{sec:discussion}
We demonstrated the potential for invertible generative models as proximal operators in unrolled architectures. In our experiments, GLOW-based proximal operators outperform several non-generative baselines, and since they learn the actual underlying structure and Probability Density Function (PDF) explicitly, they allow the user to "probe" (i.e., sample from) this PDF. Normalizing flows are particularly suited, as they allow for exact likelihood calculation, contrary to for example GANs. Note that in this work we have chosen to use the GLOW model, however, in principle any invertible flow model can be used with our method.


Compared to disjointly trained generative models used in e.g. the plug-and-play framework, unrolling the network and performing end-to-end training allows the learned PDF's to adapt to a given task. In our experiments on inpainting, we observed that directly using the pre-trained GLOW model (without any retraining to adapt to the task), yielded significantly worse results. 
This indicates that task-based adaptation of the PDF in an unrolled scheme dramatically improves performance. 

However, once we perform unrolled training, thereby adapting the PDF, we no longer promote normality for each of the latent spaces of the different flow models. While this theoretically prohibits sampling from the PDFs to probe their behaviour, we in practice noticed that naive sampling of latents (using a Gaussian) yielded structured outputs nonetheless. In future work we consider promoting such normality after each shrinkage step by including a maximum likelihood loss (i.e. \eqref{eqn:pretrainloss}) in addition to the final reconstruction loss during unrolled training.

A critical component of unrolling is the untying of the weights over the iterations. Our choice to untie the weights follows from literature, such as, Hershey et al. \cite{hershey2014}, who claim that untying the model parameters across layers in an unfolded framework helps create a more robust network. Because through untying the model parameters, we increase the network flexibility to learn a specific task while at the same time keeping the advantages of the original unfolded algorithm as being interpretable and stable. Many following works have also chosen to do so, see e.g. \cite{chien2017deep} and \cite{bertocchi2020deep}. Our own initial experiments back this up, and we observed worse performance when not untying the weights of the model.

A downside of normalizing flows is the amount of flows that need to be stacked in order to model sufficiently rich distributions, and the associated memory footprint. In order to train the unrolled architecture with multiple GLOW models end-to-end under such memory footprints, we leverage their invertible nature and reduce the memory complexity to $O(1)$ following the work of Putzky and Welling \cite{pputzky2019}. Nevertheless, the computational complexity is untouched by this approach, and normalizing flows still require more computations than the convolutional baselines we compared to in this work. 

In this work, we only experimented on fairly uni-modal datasets (Celeb-A and animefaces), that is, all the examples in the dataset are centered and cropped faces. A rich multi-modal dataset of e.g. natural images might present additional challenges for our current approach: the flows would need to map all the modes of the data distribution into a single Gaussian distribution. However, because the mapping needs to be invertible, different modes will inevitably be `connected' (i.e. there will be mass between them). It is an open question how and if the shrinkage step we perform would be able to deal with this. For example, if the shrinkage step is too large, we might move from the region of one mode in latent space to that of another mode.

In future work we also consider expanding the application space to different types of data and problems including medical imaging (e.g. MRI, CT, ultrasound), time-series (ECG, audio, wireless communication), graphs, point clouds, and compressed sensing.

\section{Conclusion}
\label{sec:conclusion}
We propose a framework for deep algorithm unfolding based on task-adapted normalizing flows priors. Our method first learns generic priors on a given training dataset, and then adapts these to the specific image restoration task at hand. We evaluate the performance of our approach for a variety of such tasks, i.e., image denoising, inpainting and deblurring and for different scenarios, i.e., out-of-distribution measurements. We demonstrate performance gains compared to strong baselines. While unfolding and end-to-end training enables fitting to (and exploiting) a specific data distribution, it also makes it more sensitive to out of distribution measurements. We show that generative flow proximal operators suffer less from this problem than standard discriminative U-Net or ResNet ones, and thus have advantages in real world applications of unfolding. Beyond image restoration, we expect our method to find applications in compressed sensing and medical image reconstruction, which is part of future work.



%



\bibliographystyle{IEEEtran} 
\bibliography{IEEEabrv, bare_jrnl}








\end{document}